\documentclass{desyproc}

\usepackage{cite, rotating, amssymb, epsfig}
\usepackage{graphics}
\begin{document}


\title{Unintegrated gluon distribution and soft $pp$ collisions at LHC
}


     \author{{\slshape Andrei Grinyuk$^1$, Hannes Jung$^{2,3}$,
     Gennady Lykasov$^1$\footnote{Speaker}, Artem Lipatov$^4$,
     Nikolay Zotov$^4$}\\[1ex]
     $^1$JINR, 141980, Dubna, Moscow region, Russia \\
     $^2$DESY, Notketra{\ss}e 85, 22607 Hamburg, Germany\\
     $^3$CERN, 1211 Gen\`eve 23, Switzerland\\
     $^4$Skobeltsyn Institute of Nuclear Physics, 
     Lomonosov Moscow State University, 119991, Moscow, Russia 
}
%
%
%
%
%
%
%
%
%



\contribID{ZZ}
\confID{UU}
\desyproc{DESY-PROC-2012-YY}
\acronym{MPI@LHC 2011}
\doi
\maketitle


\begin{abstract}
We found the parameterization of the unintegrated  gluon distribution
from the best description of the LHC data on the inclusive spectra of hadrons 
produced in $pp$ collisions at the mid-rapidity region and small transverse momenta. 
It is different from the one obtained within perturbative QCD only at low 
intrinsic transverse momenta $k_t$. The application of this distribution to analysis 
of the $e-p$ DIS allows us to get the results which do not contradict the H1 and 
ZEUS data on the structure functions at low $x$. So, the connection between the 
soft processes at LHC and low-$x$ physics at HERA is found.

\end{abstract}



\section{Introduction}
\label{1}

As is well known, hard processes involving incoming protons, such as deep-inelastic 
lepton-proton
scattering (DIS), are described using the scale-dependent parton density functions. 
Usually, these quantities are 
calculated as a function of the Bjorken variable $x$
and the square of the four-momentum transfer $q^2=-Q^2$
within the framework of popular collinear QCD factorization based on the DGLAP evolution equations
\cite{{DGLAP}}.
However, for semi-inclusive processes (such as inclusive jet production in DIS, 
electroweak boson production \cite{Ryskin:2003}, etc.) at high energies it is more appropriate
to use the parton distributions unintegrated over the transverse momentum $k_t$ in the framework 
of $k_t$-factorization QCD approach \cite{kT} 
\footnote{See, for example, reviews \cite{Andersson:02,kTreview} for more information.}. 
The $k_t$-factorization formalism is based on the BFKL \cite{BFKL} or 
CCFM \cite{CCFM} evolution equations and 
provides solid theoretical grounds for the effects of initial gluon radiation and intrinsic
parton transverse momentum $k_t$.
The theoretical analysis of the unintegrated quark $q(x,k_t)$ distribution (u.q.d.) and gluon 
$g(x,k_t)$ distribution (u.g.d.) can 
be found, for example, in \cite{GBW:98}-\cite{Nikol_Zakhar:91}.
According to  \cite{Ryskin:2010}, the u.g.d. 
$g(x,k_t)$ at fixed $Q^2$ has the very interesting behavior at small $x\leq 0.01$, it increases
very fast starting from almost zero values of $k_t$ and decreases smoothly at large $k_t$.   
In contrast, the u.q.d. $q(x,k_t)$ is almost constant 
in the whole region of $k_t$ up to $k_t\sim$ 100 GeV$/c$ and much smaller than $g(x,k_t)$. 
These distributions were obtained using the so-called KMR formalism 
within the leading order (LO) and next-to-leading order 
of QCD (NLO) 
at large $Q^2$ 
from the known
(DGLAP-evolved \cite{{DGLAP}}) parton densities determined from the global data analysis. 
The unintegrated parton distributions were succefully applied to analyze the DIS data at low $x$ and
the number of processes studied at LHC (see, for example, 
\cite{KLZ:02}-\cite{Zotov:12}.
However, at small values of $Q^2$ the nonperturbative effects should be included to 
evaluate these distributions.
The nonperturbative effects can arise from the complex structure of the QCD vacuum.  
For example, 
within the instanton approach \cite{Kochelev:1998} the very fast increase of the unintegrated 
gluon distribution function at $0\le k_t\le 0.5$ GeV$/c$ and $Q^2=1$ (GeV$/c$)$^2$ is also shown.
These results stimulated us to assume, that the unintegrated gluon distribution in the 
proton can be included by
analyzing also the soft hadron production in $pp$ collisions.  
In this paper we analyze the inclusive spectra of the hadrons produced in $pp$
collisions at LHC energies in the mid-rapidity region including the possible creation 
of soft gluons in the proton. 
We estimate the u.g.d. at low intrinsic transverse
momenta $k_t\leq 1.5-1.6$ GeV$/$c and its parameters extracted from the best description of the
LHC data at low transverse momenta $p_t$ of the produced hadrons. We also show that our u.g.d.
similar to the u.g.d. obtained in \cite{GBW:99,Jung:04} at large $k_t$
and different from it at low $k_t$. 
\section{Inclusive spectra of hadrons in $pp$ collisions}
\label{sec:1}
\subsection{Unintegrated gluon distributions}

As was mentioned above, the unintegrated gluon densities $g(x,k_t^2)$ can be generally described by 
the BFKL \cite{BFKL} evolution equation, where the leading $\ln(1/x)$ contributions are 
summed. The conventional integrated gluon distribution $g(x,Q^2)$ can be approximately related to the
unintegrated one by \cite{KLZ:02}
\begin{eqnarray}
g(x,Q^2)\sim g(x,Q_0^2)+\int_{Q_0^2}^{Q^2} dk_t^2 g(x,k_t^2) 
\label{def:BFKL}
\end{eqnarray}
where $Q_0^2$ is a starting nonzero value of $Q^2$.
An appropriate description valid for small and large $x$ is given by the CCFM \cite{CCFM} 
evolution equation that results in the u.g.d. $g(x,k_t^2,{\bar q}^2)$ 
as a function of $x, k_t^2$ and the additional scale ${\bar q}$ (see details in
\cite{Andersson:02,Jung:04} and references therein). The another u.g.d. $g(x,k_t^2)$  
satisfies the DGLAP-type evolution equation with respect to $k_t^2$ \cite{Ryskin:2003,Ryskin:2010}.  

A simple parameterization of the u.g.d. was obtained, for example, within the color-dipole approach
in \cite{GBW:98,GBW:99} on the assumption of a saturation of the gluon density at low $Q^2$ which succcefully 
described both inclusive and diffractive $e-p$ scattering. The u.g.d. $xg(x,k_t^2, Q_0^2)$ is given by
\cite{GBW:99,Jung:04}
\begin{eqnarray}
xg(x,k_t,Q_0)=
\frac{3\sigma_0}{4\pi^2\alpha_s(Q_0)}k_t^2
\exp\left(-R_0^2(x)k^2_t\right)~; 
~R_0=\frac{1}{Q_0}\left(\frac{x}{x_0}\right)^{\lambda/2}~,
\label{def:GBWgl}
\end{eqnarray}
where $\sigma_0~=~$29.12 mb, $\alpha_s~=$~0.2, $Q_0~=$~1~GeV$/$c, $\lambda~=$~0.277 and
$x_0~=$~4.1$\cdot$ 10$^{-5}$.   
The form for $xg(x,k_t,Q_0)$ given by Eq.(\ref{def:GBWgl}) was obtained in  \cite{GBW:99} within the model of
the $(q {\bar q})$ dipole \cite{Nikol_Zakhar:91,Ivan_Nikol:02} on the assumption of the saturation effect 
for the gluon density, e.g., the
dipole-nucleon cross section $\sigma_{\gamma* N}$ is assumed to be a constant at low $Q^2$ (and low $x$ too). 
It corresponds to the Gaussian form for $\sigma_{\gamma* N}({\bf b})$ as a function of the 
impact parameter ${\bf b}$.
In fact, this form could be more complicated. In this paper we study this point and try to find the 
parameterization   
for $xg(x,k_t,Q_0)$, which is related to $\sigma_{\gamma* N}({\bf b})$, from the best description of the 
inclusive spectra 
of charge hadrons produced in $pp$ collisions at LHC energies and mid-rapidity region.      
\subsection{Quark-gluon string model (QGSM) including gluons}
As is well known, the soft hadron production in $pp$ collisions at not large transfer 
can be analyzed within the soft QCD models, namely, the quark-gluon string model (QGSM)
\cite{kaid1}-\cite{BLL:2010} or the dual parton model (DPM) \cite{capell2}. The cut n-pomeron 
graphs calculated within these models result in a reasonable contribution at small but 
nonzero rapidities. However, it has been shown recently \cite{BGLP:2011} that there are some 
difficulties in using the QGSM to analyze the inclusive spectra in $pp$ collisions 
in the mid-rapidity region and at the initial energies above the ISR one. It is due to the 
Abramovskiy-Gribov-Kancheli cutting rules (AGK) \cite{AGK} at mid-rapidity ($y\simeq 0$), when 
only one-pomeron Mueller-Kancheli diagrams contribute to the inclusive spectrum 
$\rho_h(y\simeq 0, p_t)$.    
To overcome these difficulties it was assumed  in \cite{BGLP:2011} that there are soft gluons in
the proton, which are split  
into $q{\bar q}$ pairs and should vanish at the zero intrinsic transverse momentum ($k_t\sim 0$).
 The total spectrum $\rho_h(y\simeq 0, p_t)$ was split  into
two parts, the quark contribution $\rho_q(y\simeq 0, p_t)$ and the gluon one 
and their energy dependence was calculated  \cite{BGLP:2011}
\begin{eqnarray}
\rho(p_t)=\rho_q(x=0,p_t)+\rho_g(x=0,p_t)~. 
g(s/s_0)^{\Delta}{\bar\phi}_{q}(0, p_t)+
\left(g(s/s_0^{\Delta}- \sigma_{nd}\right)
{\bar\phi}_g(0, p_t)~.
\label{def:rhoagk}
\end{eqnarray}
The following parameterization for ${\tilde\phi}_q(0,p_t)$ and
 ${\tilde\phi}_g(0,p_t)$ was found \cite{BGLP:2011}:   
\begin{eqnarray}
{\tilde\phi}_q(0,p_t)=A_q\exp(-b_q p_t)~
\nonumber \\
{\tilde\phi}_g(0,p_t)=A_g\sqrt{p_t}\exp(-b_g p_t),
\label{def:phiq}
\end{eqnarray}
where $s_0=1 GeV^2, g=21 mb, \Delta=0.12$.
The parameters are fixed  from the fit to the data on the $p_t$ distribution of
charged particles at $y=0$ \cite{BGLP:2011}:
$A_q=4.78\pm 0.16$ (GeV$/$c)$^{-2}$,~$b_q=7.24\pm 0.11$ (Gev/c)$^{-1}$ and
 $A_g=1.42\pm 0.05$ (GeV$/$c)$^{-2}$;~ 
$b_g=3.46\pm 0.02$ (GeV/c)$^{-1}$. 
\begin{figure}[h!!]
\centerline{\includegraphics[width=0.4\textwidth]{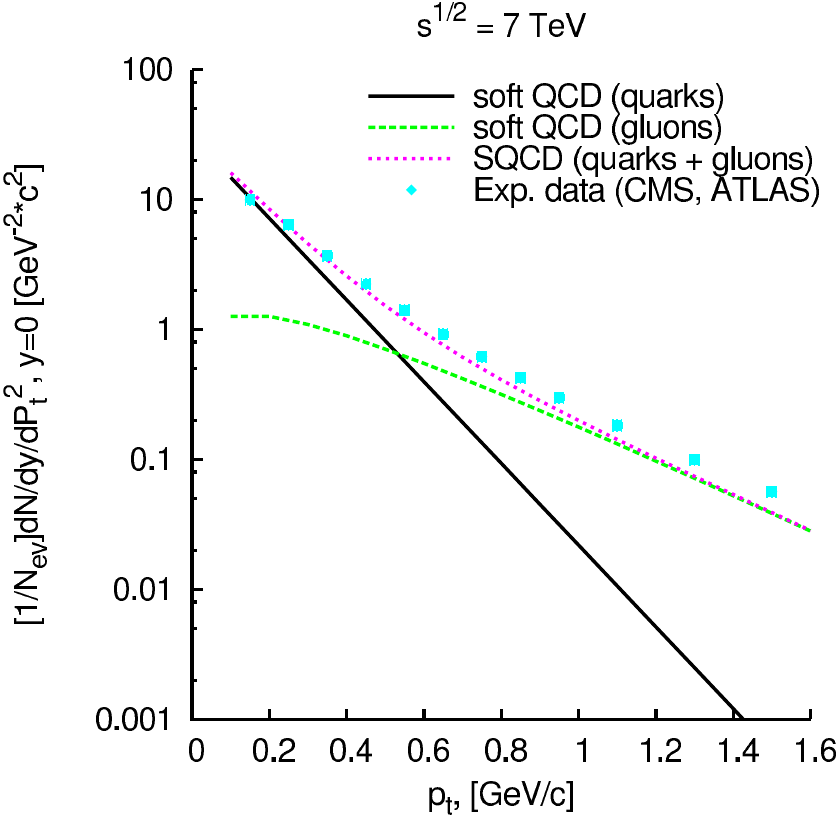}}
  \caption[Fig.2]{The inclusive spectrum of the charged hadron as a function of $p_t$ (GeV$/c$)
in the central rapidity region ($y=0$) at $\sqrt{s}=$7 TeV 
at $p_t\leq$ 1.6 GeV$/$c compared with the CMS \cite{CMS} which are very close to the
ATLAS data \cite{ATLAS}.}
\end{figure} 
Figure 1 illustrates the best fit of the inclusive spectrum of charged hadrons produced in 
$pp$ collisions at $\sqrt{s}=$7 TeV and the central  rapidity region at the hadron transverse momenta
$p_t\leq$ 1.6 Gev$/$c; the solid line corresponds to the quark contribution $\rho_q$,  
the dashed line is the gluon contribution $\rho_g$, and the dotted curve is the sum of these contributions
$\rho_h$ given by Eq.(\ref{def:rhoagk}).
\subsection{Modified unintegrated gluon distributions}
We calculated the gluon contribution  ${\tilde\phi}_g(0,p_t)$ entering into 
Eq.(\ref{def:phiq}) as 
the cut graph of the one-pomeron exchange in the gluon-gluon interaction (Fig.2, right) 
using the splitting of the gluons into the $q{\bar q}$ pair. Then the calculation was 
made in a way similar to the calculation of the sea quark contribution to the inclusive 
spectrum within the QGSM, see Eqs.(4,5) in \cite{BGLP:2011} at $n=2$.
\begin{figure}[h!]
\centerline{\includegraphics[width=0.8\textwidth]{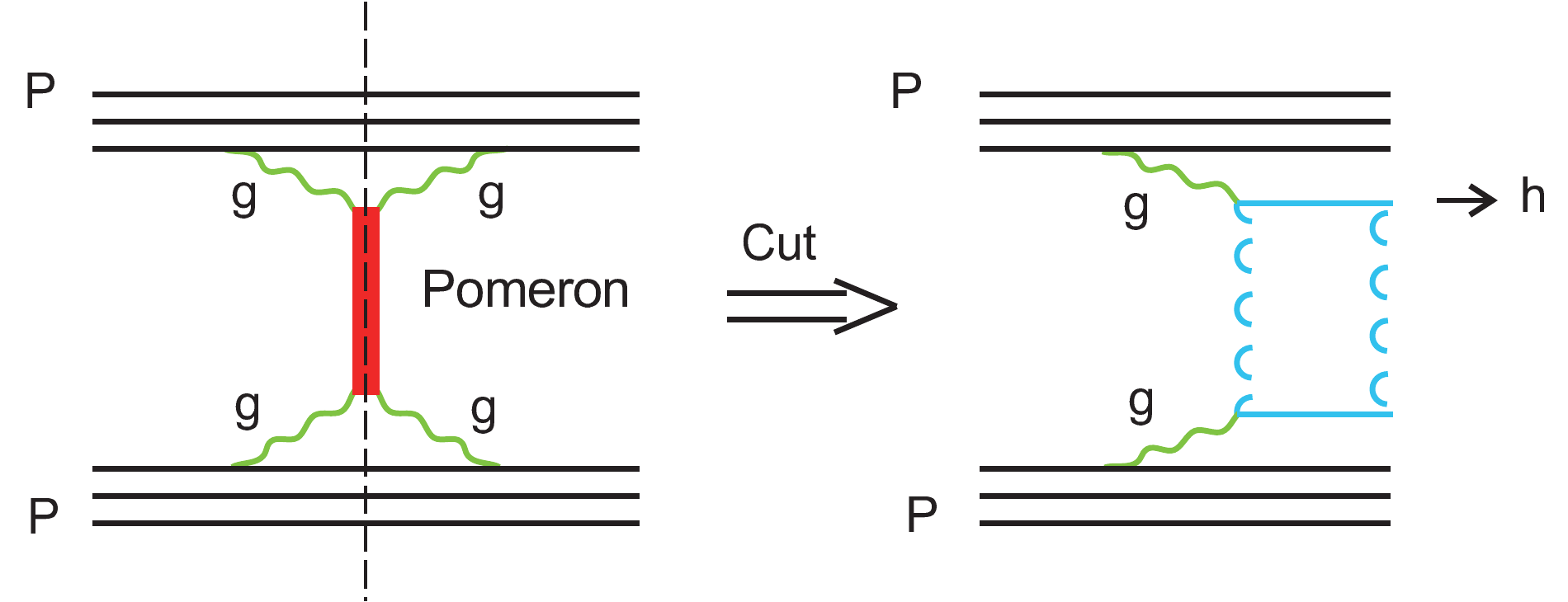}}
\caption 
    {The one-pomeron exchange graph between two gluons in the elastic $pp$ scattering (left) 
and the corresponding cut graph (right).} 
\label{Fig_1}
\end{figure} 
\begin{eqnarray}
\rho_g(x_\pm, p_{ht})=F(x_+,p_{ht})F(x_-,p_{ht})~,
\label{def:rhog}
\end{eqnarray}
where
\begin{eqnarray}
F_(x_\pm,p_t;p_{ht})=
\int_{x\pm}^1dx_1
\int d^2k_{1t}f_{q,{\bar q}}(x_1,k_{1t})G_{q({\bar q})\rightarrow h}
\left(\frac{x_\pm}{x_1},p_{ht}-k_t)\right)~,
\label{def:Fqbrq}
\end{eqnarray}
Here $G_{q({\bar q})\rightarrow h}(z,{\tilde k}_t)=zD_{q({\bar q})\rightarrow h}(z,{\tilde k}_t)$,
$D_{q({\bar q})\rightarrow h}(z,{\tilde k}_t)$ is the fragmentation function (FF) of the quark (antiquark)
to a hadron $h$, $z=x_\pm/x_1,{\tilde k}_t=p_{ht}-k_t$, 
$x_{\pm}=0.5(\sqrt{x^2+x_t^2}\pm x), x_t=2\sqrt{(m_h^2+p_t^2)/s}$.  
The distribution of sea quarks (antiquark) 
$f_{q,{\bar q}}$ is related to the splitting function ${\cal P}_{g\rightarrow q{\bar q}}$ of gluons to 
$q{\bar q}$ by
 \begin{eqnarray}
f_{q,{\bar q}}(z,k_t)=\int_z^1 g(z_1,k_t,Q_0){\cal P}_{g\rightarrow q{\bar q}}
(\frac{z}{z_1})~,
\label{def:fqbq}
\end{eqnarray}
where $g(z_1,k_{1t},Q_0)$ is the u.g.d.. The gluon splitting function ${\cal P}_{g\rightarrow q{\bar q}}$
was calculated within the Born approximation. 

Calculating the diagram of Fig.2 (right) by the use of Eqs.(\ref{def:rhog}-\ref{def:fqbq}) for the gluon 
contribution $\rho_g$ we took the FF to charged hadrons, pions, kaons, and $p{\bar p}$ pairs obtained
within the QGSM \cite{Shabelsk:1992}. From the best description of $\rho_g(x\simeq 0,p_{ht}$, see
its parameterization given by Eq.(\ref{def:phiq}), we found the form     
for the $xg(x,k_t,Q_0)$ which was fitted in the following form:
\begin{eqnarray}
xg(x,k_t,Q_0)=\frac{3\sigma_0}{4\pi^2\alpha_s(Q_0)} 
C_1 (1-x)^{b_g}\times
\nonumber \\
\left(R_0^2(x)k_t^2+C_2(R_0(x)k_t)^a\right)
\exp\left(-R_0(x)k_t-d(R_0(x)k_t)^3\right)~,
\label{def:gldistrnew}
\end{eqnarray}
The coefficient $C_1$ was found from the following normalization:
\begin{eqnarray}
g(x,Q_0^2)=\int_0^{Q_0^2} dk_t^2g(x,k_t^2,Q_0^2)~,
\label{def:BFKL}
\end{eqnarray}
and the parameters 
$$
a=0.7; C_2\simeq 2.3; \lambda=0.22; b_g=12; d=0.2; C_3=0.3295
$$
were found from the best fit of the LHC data on the inclusive spectrum of charged
hadrons produced in $pp$ collisions and in the mid-rapidity region at $p_t\leq$1.6 GeV$/$c,
see the dashed line in Fig.1 and Eq.(\ref{def:phiq}).

\begin{figure}[h!]
\centerline{\includegraphics[width=0.8\textwidth]{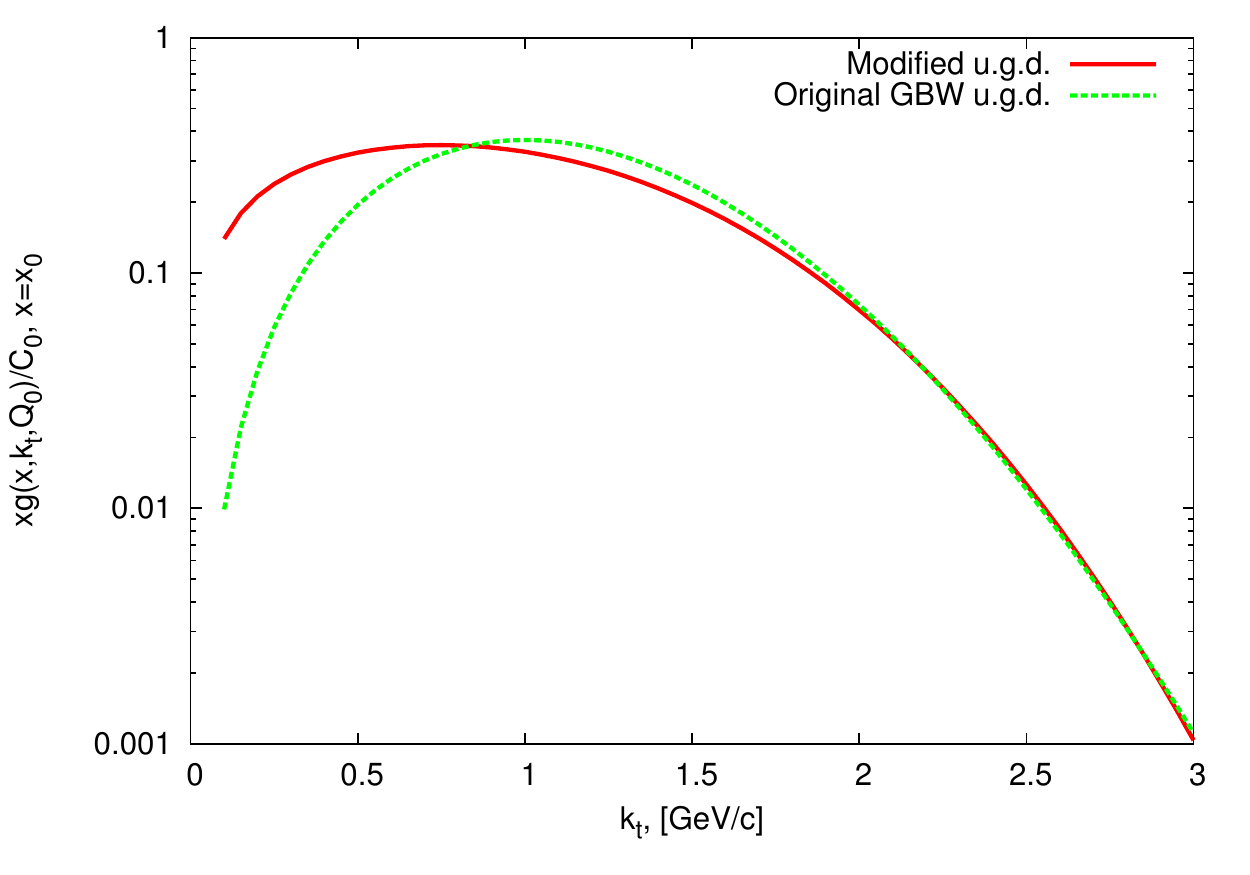}}
\caption 
    {The unintegrated  gluon distribution $xg(x,k_t,Q_0)/C_0$ as a function of $k_t$
at $x=x_0$ and $Q_0=1.$GeV$/$c. The dashed curve corresponds to the original GBW 
\cite{GBW:99,Jung:04}, Eq.(\ref{def:GBWgl}), and the solid line is the modified u.g.d. 
given by Eq.(\ref{def:gldistrnew}).
} 
\label{Fig_3}
\end{figure} 
Figure 3 presents the modified u.g.d. obtained by calculating the cut one-pomeron graph of Fig.2
and the original GBW u.g.d. \cite{GBW:99,Jung:04} as a function of the transverse gluon momentum
$k_t$. Here $C_0=3\sigma_0/(4\pi^2\alpha_s(Q_0))$. One can see that the modified u.g.d.
(the solid line in Fig.3) is different from the original GBW u.g.d. \cite{GBW:99,Jung:04} at $k_t~<~1.5$ GeV$/$c
and coincides with it at larger $k_t$. This is due to the sizable contribution of $\rho_g$ 
(Eqs.(\ref{def:rhoagk},\ref{def:phiq})) to the inclusive spectrum $\rho(p_t)$ of charged hadrons produced
in $pp$ collisions at LHC energies and in the mid-rapidity region, see the dashed line in Fig.1. 

\section{Proton longitudinal structure function}

Within the $k_t$-factorization approach, 
the proton longitudinal structure function $F_{L}(x,Q^2)$ calculated
in the leading order of QCD can be presented in the following form \cite{KLZ:02}: 
\begin{eqnarray}
F_{L}(x,Q^2)=\int_x^1\frac{dz}{z}\int^{Q^2} dk_t^2\sum_{flavour (f)}e_f^2 \,{\hat {\cal C}}^g_{L}
(\frac{x}{z},Q^2,m_f^2,k_t^2)g(x,k_t^2)~,
\label{def:F2L}
\end{eqnarray}
where $e_f^2$ is the charge of the quarks of flavor $f$, the functions 
${\hat {\cal C}}^g_{2,L}(x/z,Q^2,m_f^2,k_t^2)$ are the so called hard structure functions of the off-shell
gluons with virtuality $k_t^2$ which correspond to the quark-box diagram for the photon-gluon
interactions  \cite{KLZ:02}.
In the present note we calculated $F_{L}(x,Q^2)$ at the fixed value of the missing mass $W = 276$ GeV 
using the parameterization for u.g.d. at fixed $Q_0^2$ given by Eq.(\ref{def:GBWgl}) 
and Eq.(\ref{def:gldistrnew}). The results of our calculations are presented in Fig.~4 in 
comparison with the H1 data \cite{H1,H1EL}.

\begin{figure}[h!]
\centerline{\includegraphics[width=0.8\textwidth]{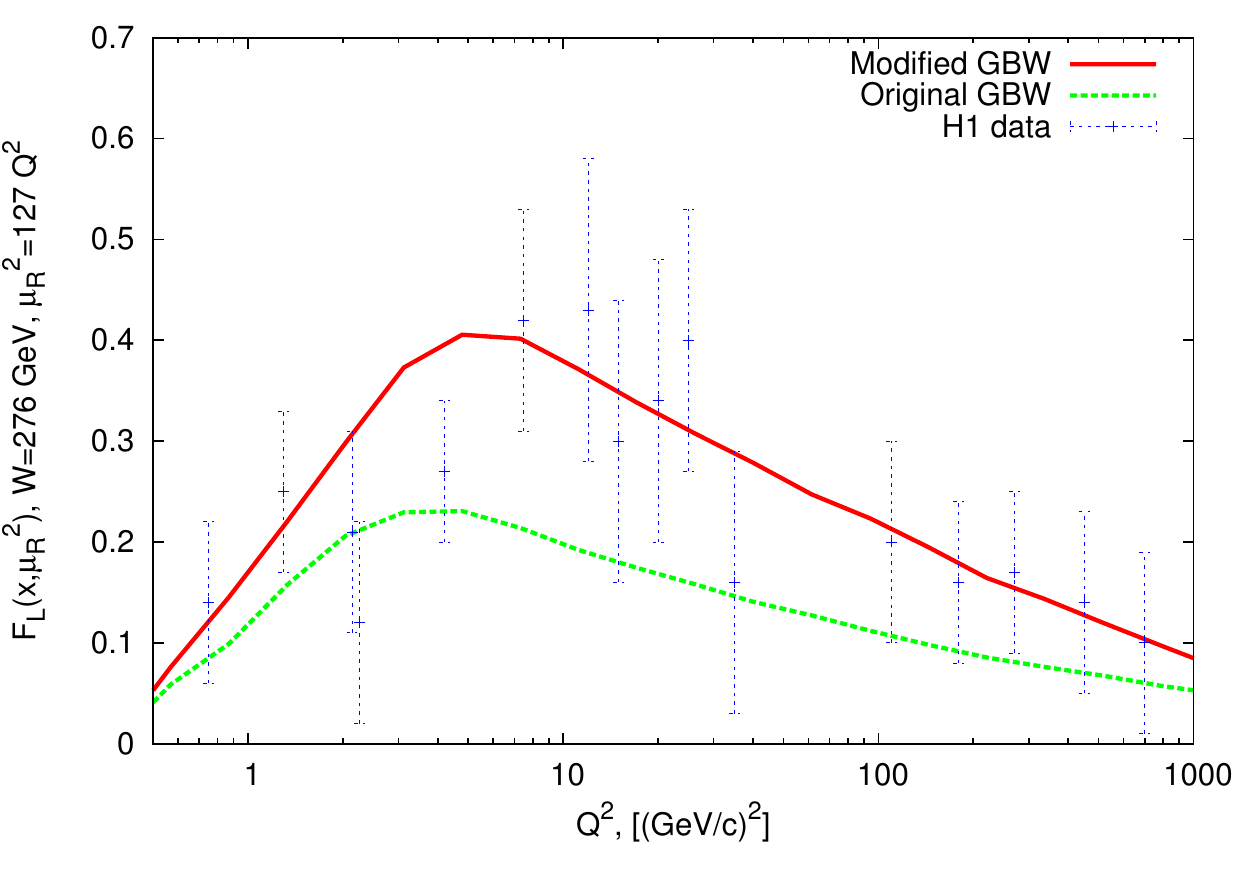}}
\caption 
{The longitudinal structure function $F_{L}(\mu_R^2)$ at $W=$276 GeV and $\mu_R^2=K\cdot Q^2$,
where $K=127$ \cite{Brodsk_K:99}.  The H1 data are taken from \cite{H1,H1EL}.     
} 
\label{Fig_4}
\end{figure} 

We find that the modified u.g.d. allows us also to describe the proton $F_L$ 
Note that in order to take into account the NLO corrections
(which are important at low $Q^2$) in our numerical calculations we apply the method 
proposed in \cite{KLZ:04}. Following \cite{KLZ:04,Brodsk_K:99}, we 
use the shifted value of the renormalization scale $\mu_R^2 = K \, Q^2$, where
$K = 127$. As is was shown in \cite{Brodsk_K:99}, this shifted scale in the DGLAP approach 
at LO approximation leads to the results which are very close to the NLO ones. 

\section{Conclusion} 

We fitted the experimental data on the inclusive
spectra of charged particles produced in the central $pp$ collisions at energies larger 
than the ISR starting
with the sum of the quark contribution $\rho_q$  and the gluon
contribution $\rho_g$ (see Eqs.(\ref{def:rhoagk},\ref{def:phiq})).
 The parameters of this fit do not depend on the initial energy in that energy interval.
Assuming creation of soft gluons in the proton at low transverse momenta $k_t$ and 
calculating the cut one-pomeron graph between two gluons in colliding protons we found
the form for the unintegrated gluon distribution (modified u.g.d) as a function of $x$ 
and $k_t$ at the fixed value of $Q_0^2$.
The parameters of this u.g.d. were found from the best description of the LHC data on 
the inclusive spectra of the charged hadrons
produced in the mid-rapidity pp collisions at low $p_t$. It was shown that the modified u.g.d.
is different from the original GBW u.g.d. obtained in  \cite{GBW:99,Jung:04}
at $k_t\leq$ 1.6 GeV$/$c and it coincides with the GBW u.g.d. at $k_t>1.6$GeV$/$c.
It was also shown that the  modified u.g.d. allows us to describe the longitudinal structure function
$F_{L}(Q^2)$ at the fixed missing mass $W$ better than the original GBW u.g.d. 
Therefore, some link between soft processes at the LHC and low-$x$ physics at HERA is found.
\vspace{0.2cm}

{\bf Acknowledgments}\\
The authors are grateful to S.P. Baranov, B.N. Ermolaev, E.A. Kuraev, L.N. Lipatov,  
M. Mangano, C. Merino, M.G. Ryskin and Yu.M. Shabelskiy for useful discussions 
and comments. A.V.L. and N.P.Z. are very grateful to the
DESY Directorate for the support within the Moscow --- DESY project on Monte-Carlo
implementation for HERA --- LHC.
A.V.L. was supported in part by the grant of the President of the
Russian Federation (MK-3977.2011.2).
This research was also supported by the 
FASI of the Russian Federation (grant NS-3920.2012.2),
FASI state contract 02.740.11.0244, 
RFBR grants 11-02-01454-a and 11-02-01538-a,
and the RMES (grant the Scientific Research on High Energy Physics).

\begin{footnotesize}

\end{footnotesize}

\end{document}